\documentclass[namedreferences]{solarphysics}
\usepackage[optionalrh]{spr-sola-addons} 
\usepackage{graphicx}        
\usepackage{color}           
\usepackage{url}             
\usepackage[pdfborder={0 0 0 },colorlinks={true}, urlcolor=blue,breaklinks]{hyperref}
\ifx \doiurl    \undefined \def \doiurl#1{\href{http://dx.doi.org/#1}{\textsf{#1}}}\fi
\ifx \adsurl    \undefined \def \adsurl#1{\href{http://adsabs.harvard.edu/abs/#1}{\textsf{#1}}}\fi
\ifx \arxivurl  \undefined \def \arxivurl#1{\href{http://arxiv.org/abs/#1}{\textsf{#1}}}\fi
\ifx \urlurl    \undefined \def \urlurl#1{\href{http://#1}{\textsf{#1}}}\fi

\newcommand{\etal}{{\it et al.}}

\newcommand{\fig}[1]{Figure~\ref{#1}}

\newcommand{\speed}[1]{#1 km~s${}^{-1}$}

\newcommand{\acc}[1]{#1 m~s${}^{-2}$}
\newcommand{\rsun}[1]{$\rm {#1}\,R_\odot$}

\newcommand{\aap}{    {\it Astron. Astrophys.}}

\newcommand{\apj}{    {\it Astrophys. J.}}
\newcommand{\apjl}{   {\it Astrophys. J. Lett.}}

\newcommand{\jgr}{    {\it J. Geophys. Res.}}
\newcommand{\mnras}{  {\it Mon. Not. Roy. Astron. Soc.}}
\newcommand{\nat}{    {\it Nature}}

\newcommand{\pasj}{   {\it Pub. Astron. Soc. Japan}}

\newcommand{\solphys}{{\it Solar Phys.}}
 
\newcommand{\ssr}{    {\it Space Sci. Rev.}}

\begin{document}

\begin{article}

\begin{opening}

\title{Observations of a Quasi-Periodic, Fast-Propagating Magnetosonic Wave in Multiwavelengths and its Interaction with Other Magnetic Structures}

\author{Y.-D~\surname{Shen}$^{1,2,3}$\sep
         Y.~\surname{Liu}$^{1,2}$\sep
        J.-T.~\surname{Su}$^{2}$\sep
        H.~\surname{Li}$^{3}$\sep
        X.-F.~\surname{Zhang}$^{1}$\sep
        Z.-J.~\surname{Tian}$^{1}$\sep
        R.-J.~\surname{Zhao}$^{1}$\sep
        A.~\surname{Elmhamdi}$^{1,4}$      
       }
\runningauthor{Y.-D. Shen {\it et al.}}
\runningtitle{Observations of a QFP wave Interacting with a Coronal Loop System}

   \institute{$^{1}$ Yunnan Astronomical Observatory, CAS, Kunming 650011, China
                     email: \href{mailto:ydshen@ynao.ac.cn}{ydshen@ynao.ac.cn}\\ 
              $^{2}$ Key Laboratory of Solar Activity, National Astronomical Observatories, CAS, Beijing 100012, China\\
              $^{3}$ Key Laboratory of Dark Matter and Space Astronomy, Purple Mountain Observatory, CAS, Nanjing 210008, China\\
              $^{4}$ Physics and Astronomy Department, College of Science, King Saud University, P.O. Box 2455, Riyadh 11451, Saudi Arabia\\}

\begin{abstract}
We present observations of a quasi-periodic fast-propagating (QFP) magnetosonic wave on 23 April 2012, with high-resolution observations taken by the {\it Atmospheric Imaging Assembly} onboard the {\it Solar Dynamics Observatory}. Three minutes after the start of a C2.0 flare, wave trains were first observed along an open divergent loop system in 171 \AA\ observations at a distance of 150 Mm from the footpoint of the guiding loop system and with a speed of \speed{689}, then they appeared in 193 \AA\ observations after their interaction with a perpendicular, underlaying loop system on the path; in the meantime; their speed decelerated to \speed{343} within a short time. The sudden deceleration of the wave trains and their appearance in 193 \AA\ observations are interpreted through a geometric effect and the density increase of the guiding loop system, respectively. We find that the wave trains have a common period of 80 seconds with the flare. In addition, a few low frequencies are also identified in the QFP wave. We propose that the generation of the period of 80 seconds was caused by the periodic releasing of energy bursts through some nonlinear processes in magnetic reconnection, while the low frequencies were possibly the leakage of pressure-driven oscillations from the photosphere or chromosphere, which could be an important source for driving coronal QFP waves. Our  results also indicate that the properties of the guiding magnetic structure, such as the distributions of magnetic field and density as well as geometry, are crucial for modulating the propagation behaviors of QFP waves.
\end{abstract}
\end{opening}

\section{Introduction}
Investigations of magnetohydrodynamic (MHD) waves in the magnetically dominated solar atmosphere have a long history. However, due to the lack of actual observations in the past, the investigations were mainly limited to theoretical studies ({\it e.g.} \opencite{robe83}, \citeyear{robe84}; \opencite{appe86}; \opencite{edwi83}, \citeyear{edwi88}), besides a few observational studies based on ground-based radio or optical telescopes ({\it e.g.} \opencite{park69}; \opencite{kout83}). In the last two decades, the launch of a series of space-borne solar telescopes such as SOHO, TRACE, STEREO, and Hinode has led to a revolutionary breakthrough in the observational study of MHD waves. However, these instruments have their own limitations for observing fast magnetosonic waves (see \inlinecite{naka05} for details). Thanks to the launch of the {\it Solar Dynamics Observatory} (SDO, \inlinecite{pesn12}) in 2010, many instrumental deficiencies are largely overcome due to the high temporal and spatial resolution and full-disk observation capability of this telescope. Previous studies have indicated that MHD waves play an important role in the context of the enigmatic problems of coronal heating and acceleration of the fast solar wind, since they can carry magnetic energy over a large distance ({\it e.g.} \opencite{scha49}; \opencite{oste61}; \opencite{wals03}; \opencite{tian11}; \opencite{mort12a}). Furthermore, MHD waves can also be used to diagnose many physical parameters of the solar corona with the so-called coronal seismology technique (\opencite{uchi70}; \opencite{robe84}). For example, with some measurable physical parameters, one can estimate the coronal magnetic-field strength (\opencite{naka01}; \opencite{west11}; \opencite{shen12b}, \citeyear{shen12c}), coronal dissipative coefficients (\opencite{naka99}), and coronal sub-resolution structures (\opencite{robb01}; \opencite{king03}; \opencite{mort12b}). These parameters are difficult to obtain with direct measurements, but they are crucial for understanding a number of complex physical processes in the solar corona.

It is generally known that there are three types of MHD waves in the solar corona, namely Alfv$\rm \acute{e}$n and slow and fast magnetosonic waves. Except for the slow-mode waves, up to the present, reports on Alfv$\rm \acute{e}$n and fast-mode waves are very rare. This is mainly due to the instrumental limitations such as low cadence. For observational investigations on quasi-periodic fast-mode waves, \inlinecite{will02} first reported a quasi-periodic fast wave that travels through the apex of an active-region coronal loop with a speed of \speed{2100} and a dominant period of six seconds. This event was observed during the total solar eclipse on 11 August 1999, with the {\it Solar Eclipse Corona Imaging System} (SECIS) instrument, which has a rapid cadence of $2.25 \times 10^{-2}$ seconds and a pixel size of $4.07^{''}$ \cite{will01}. This temporal resolution is sufficient to detect the short-period fast waves. In an open magnetic-field structure, \inlinecite{verw05} found fast-propagating transverse waves that have phase speeds in the ranges of \speed{200--700} and periods in the range of 90--220 seconds. The authors interpreted them as propagating fast magnetosonic kink waves guided by a vertical, evolving, open structure. Solar decimetric radio emission of fiber bursts are often interpreted as a signature of magnetosonic wave trains in the solar corona. They often have a period of minutes and show a ``tadpole'' structure in the wavelet spectra ({\it e.g.} \inlinecite{mesz09a}, \citeyear{mesz09b}, \citeyear{mesz13}; \inlinecite{mesz11}; \inlinecite{karl11}), as predicted in theory studies ({\it e.g.} \inlinecite{naka04}; \inlinecite{jeli12}). 

With the high temporal and spatial resolution observations of  the {\it Atmospheric Imaging Assembly} (AIA, \inlinecite{leme12}; \inlinecite{boer12}) instrument onboard SDO, a new type of MHD wave dubbed quasi-periodic fast-propagating magnetosonic waves (QFP) has been detected recently. Such waves have multiple arc-shaped wave trains, and they are often observed in diffuse open coronal loops at 171 \AA\ temperatures (Fe {\it \uppercase\expandafter{\romannumeral 9}}; $\log T=5.8$). Initial observational results indicate that QFP waves have an intimate relationship with the accompanying flare. However, questions about their generation, propagation, and energy dissipation are still open questions. \inlinecite{liu11} presented the first QFP wave study with observations taken by SDO/AIA, and they found that multiple arc-shaped wave trains successively emanate from near the flare kernel and propagate outward along a funnel-like structure of coronal loops with a phase speed of about \speed{2200}. With Fourier analysis, they detected three dominant frequencies of 5.5, 14.5, and 25.1 mHz in the QFP wave, in which the frequency of 5.5 mHz temporally coincides with quasi-periodic pulsations of the accompanying flare, which suggests that the flare and the QFP wave were possibly excited by a common origin. \inlinecite{shen12b} investigated a similar case that occurred on 30 May 2011, and they compared the frequencies of the QFP wave and the accompanying flare. Their observational results indicate that all of the flare's frequencies can be found in the wave's frequency spectrum, but a few low frequencies of the QFP wave are not consistent with those of the flare. Thus they proposed that the leakage of pressure-driven oscillations from photosphere into the low corona could be another source for driving QFP waves. Recently, \inlinecite{yuan13} reanalyzed the event on 30 May 2011 with AIA data and radio observations provided by the Nancay Radioheliograph. They found that the QFP wave could be divided into three distinct sub-QFP waves that have different amplitudes, speeds, and wavelengths. In addition, the radio emission show three radio bursts that are highly correlated in start time with the sub-QFP waves. This result suggests that the generation of QFP waves should be tightly related with the regimes of energy releasing in magnetic reconnections. QFP waves coupling with diffuse single broad pulse of extreme-ultraviolet (EUV) waves (so-called ``EIT waves" ({\it e.g.} \opencite{thom98}; \opencite{shen12a})) are observed recently by \inlinecite{liu12}. The authors found that multiple wave trains propagate ahead of and behind of a coronal mass ejection (CME) simultaneously. However, the two components of the wave trains have different speeds and periods, in which only those running ahead of the CME have similar period to the flare. Modeling efforts have been made to understand the physics in QFP waves \cite{naka03,bogd03,hegg09,fedu11,ofma11}. Especially, \inlinecite{ofma11} performed a three-dimensional numerical simulation for the QFP wave presented by \inlinecite{liu11}. They successfully reproduced the multiple arc-shaped wave trains that have similar amplitude, wavelength, and propagation speeds as those obtained from observation.

In this article, we present an observational study of a QFP wave that occurred on 23 April 2012 and was accompanied by a {\it Geostationary Operational Environmental Satellite} (GOES) C2.0 flare in NOAA active region AR11461 (N12, W20). The wave trains were first observed in 171 \AA\ observations; however, after their interaction with another loop system on the path, they appeared in the hotter 193 \AA\ observations. In the meantime, the speed of the wave trains decelerated to about half of that before the interaction. With the Fourier and wavelet analysis techniques, we study the periodicity, generation, and propagation of the QFP wave, then possible mechanisms for the quick deceleration of the wave trains during the interaction and their sudden appearance in 193 \AA\ observations are discussed. 

\section{Observations}
AIA  onboard {\em SDO} is very suitable for detecting fast-propagating features such as fast magnetosonic waves with short periods. It captures images of the Sun's atmosphere out to \rsun{1.3} and has high temporal resolution of as short as 12 seconds. AIA produces imaging data with four $4096 \times 4096$ detectors with a pixel size of $0.6^{"}$, corresponding to an effective spatial resolution of $1.2^{"}$ in seven EUV and three UV-visible channels, which cover a wide temperature range from $\log T=3.7$ to $\log T=7.3$. All of these parameters are necessary ingredients for detecting fast-propagating waves. In the presented case, the wave trains were firstly captured in AIA 171 \AA\ (Fe {\it \uppercase\expandafter{\romannumeral 9}}; $\log T=5.8$) and then in 193 \AA\ (Fe {\it \uppercase\expandafter{\romannumeral 12}}, {\it \uppercase\expandafter{\romannumeral 24}}; $\log T = 6.2, 7.3$) observations. We study the QFP wave using the running-difference, base, and running-ratio images, in which the running-difference images are constructed by subtracting each image with the previous one, the base-ratio images are obtained by dividing the time-sequence images with a pre-event image, and the running-ratio images are obtained by dividing each image with the previous one. In addition, the GOES soft X-ray fluxes are also used to analyze the periodicity of the accompanying C2.0 flare. The AIA images used in this article are calibrated with the standard procedure aia\_prep.pro available in SolarSoftWare (SSW) and then differentially rotated to a reference time (17:30:00 UT), and solar North is up, West to the right.
  
\section{Results}
\subsection{Overview of the QFP Wave on 23 April 2012}
The QFP wave on 23 April 2012 was accompanied by a GOES C2.0 flare (N13, W17) in NOAA AR11461 (N12, W20), a global EUV wave, and a coronal mass ejection (CME). According to the GOES flare record, the start, peak, and end times of the flare are 17:37, 17:51, and 18:05 UT, respectively. The QFP wave could be observed about three minutes after the flare start, which indicates that the generation of the flare and the wave trains may have some internal physical relations. On the other hand, the relationship between the QFP wave and the preceding EUV wave is not obvious. Therefore, we will confine our attention to the QFP wave and the accompanying flare in the present article. Detailed analysis of the global EUV wave has been published very recently by \inlinecite{shen13}.

\begin{figure}    
\centerline{\includegraphics[width=0.95\textwidth,clip=]{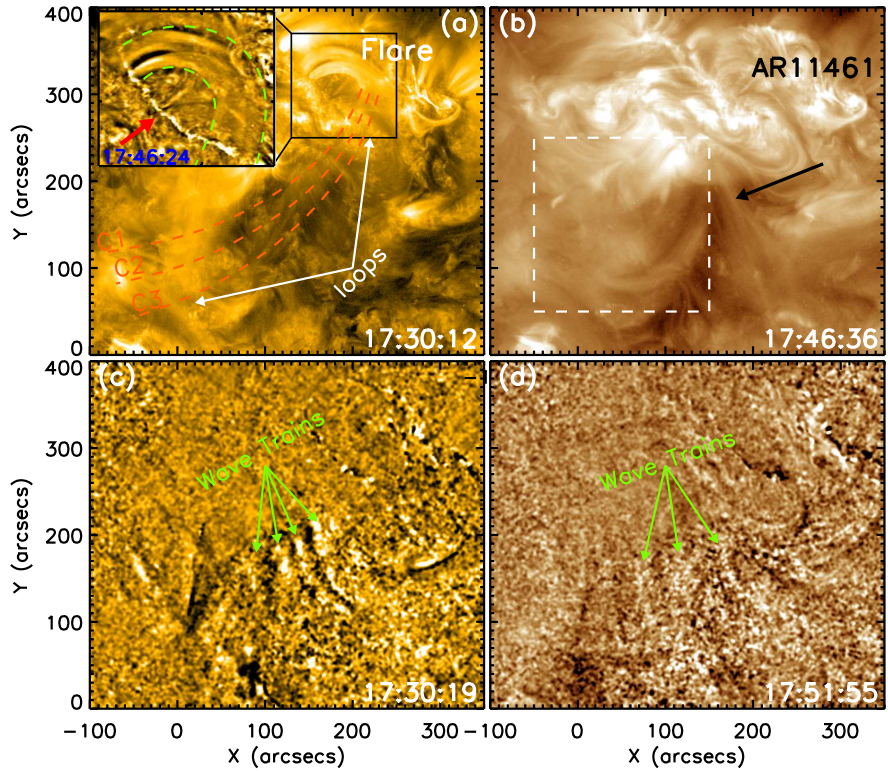}}
\caption{An overview of the QFP wave on 23 April 2012. (a) AIA 171 \AA\ and (b) AIA 193 \AA\ raw images show the pre-event magnetic environment, while (c) AIA 171 \AA\ and (d) AIA 193 \AA\ are base-ratio images displaying the multiple wave trains. The three orange red dashed curves in panel (a) are used to obtain the time--distance diagrams shown in Figure \ref{fig2}, and the guiding loop is indicated by the two white arrows. The inset in panel (a) is a close-up view of the black box region at 17:46:24 UT. It is a filtered image obtained by subtracting a smoothed image with a boxcar average over $15 \times 15$ pixels. In the inset, the long flare ribbon is indicated by the red arrow, and the two green dashed curves outline the loop system that guides the wave trains. In panel (b), the white dashed box indicates the region where Fourier analysis is applied, while the black arrow points to the perpendicular loop system. The arrows in panels (c) and (d) point to the multiple wave trains. The field of view is $450^{''} \times 400^{''}$ for each frame and an animation for this figure is available in the electronic supplementary materials.}
\label{fig1}
\end{figure}

The wave trains were primarily observed in the 171 \AA\ observations along an open loop system rooted in active region AR11461. Furthermore, the wave trains were also observed in the 193 \AA\ observations after a few minutes. This phenomenon is different from the cases that have been documented in previous studies, where wave trains can only be identified at the 171 \AA\ temperature \cite{liu11,shen12b}. The pre-event magnetic condition of the source region and the morphology of the wave trains are displayed in \fig{fig1}. It can be seen that the path of the wave trains was along the diverging coronal loop system, which can be identified in the 171 \AA\ raw image as indicated by the white arrows (see \fig{fig1}(a) and the animation available in the electronic supplementary material). On the path of the wave trains, there is another loop system that was nearly perpendicular to the guiding field of the wave trains (see the black arrows in \fig{fig1}(b) and the animation). The propagation of the wave trains were inevitably influenced by this perpendicular loop system, which will be analyzed in detail using time--distance diagrams obtained from the red dashed curves as shown in \fig{fig1}(a). In \fig{fig1}(c) and (d), we show the multiple arc-shaped wave trains in running-ratio 171 \AA\ (\fig{fig1}(c)) and 193 \AA\ (\fig{fig1}(d)) images. They emanated successively from the footpoint of the guiding loop and faded in sequence at a distance of about 300 Mm from the guiding loop's footpoint. The successive wave trains were manifested as alternating white-black-white fringes. The footpoint region of the guiding loop system is highlighted in the small inset in \fig{fig1}(a), in which the loop system is outlined using two dashed-green curves. It is interesting that a long flare ribbon lay close to the footpoint of the guiding loop system. In consideration of the temporal  relationship between the flare and the QFP wave, we conjecture that this flare ribbon might be a direct evidence for the generation of the wave trains. However, the wave trains did not show up immediately following the appearance of the flare ribbon, but rather appeared at a distance of about 150 Mm from the flare ribbon (also the footpoint of the guiding loop system) in the 171 \AA\ images. Here the distance is measured along the curving coronal loop rather than a straight-line distance. As a comparison, the distance is about 260 Mm from the flare ribbon when the wave trains could be observed in the 193 \AA\ images. From the time-sequence observations, we determine that the lifetimes of the wave trains are about 15 (17:40 -- 17:55 UT) and 8 (17:47 -- 17:55 UT) minutes at 171 \AA\ and 193 \AA\ wavelength bands, respectively. The start time of  wave trains in the 171 (193) \AA\ observations is delayed relative to that of the flare by about three (ten) minutes, while the appearance time in the 193 \AA\ images is delayed relative to that from 171 \AA\ by about seven minutes.

\subsection{Kinematics Analysis of the Wave Trains}
\begin{figure}    
\centerline{\includegraphics[width=0.95\textwidth,clip=]{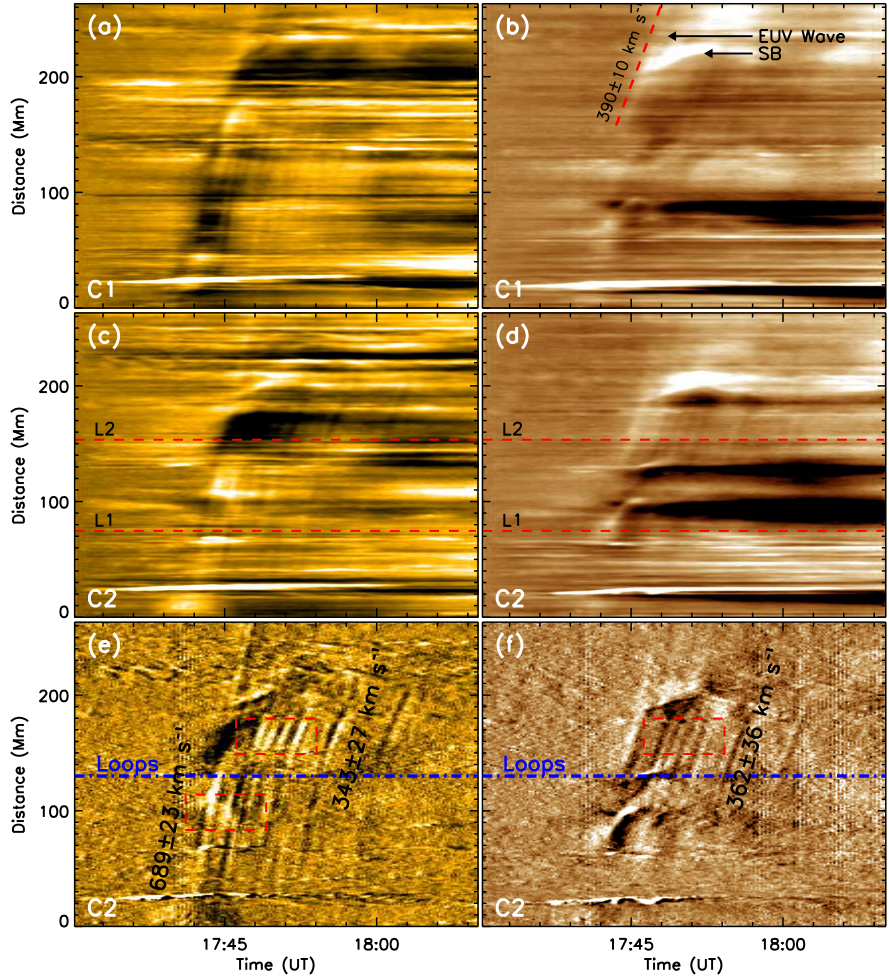}}
\caption{time--distance diagrams show the kinematics of the wave trains. The top and middle rows are obtained from base-ratio images, while the bottom row are made from running-ratio images. The left and right columns are obtained from 171 and 193 \AA\ images respectively. The two black arrows in panel (b) point to the EUV wave stripe and the stationary brightening. The red dashed lines in panels (c) and (d) mark the positions where we analyze the periodicity of the wave trains, while the blue dash--dot line in panels (e) and (f) indicate the position of the perpendicular loop system. The average speeds of the EUV wave and the wave trains are also plotted in the figure. The three dashed red boxes mark the regions shown in \fig{fig3}, in which the top one in panel (e) and the one in panel (f) indicate the same region.}
\label{fig2}
\end{figure}

We study the kinematics of the wave trains using time--distance diagrams obtained along curves perpendicular to the propagation direction of the wave trains (see \fig{fig1}(a)). To make a time--distance diagram, we first obtain the intensity profiles along a curve from time sequence images by averaging ten pixels across the curve. Then, a time--distance diagram can be created by stacking the obtained profiles in time sequence. \fig{fig2} shows the time--distance diagrams made from base- and running-ratio 171 \AA\ and 193 \AA\ observations along cuts C1 and C2. The base-ratio time--distance diagrams show best the broad EUV wave stripe and dark dimming regions that are thought to be an effect of density decrease rather than temperature change \cite{jian07,shen10}, while the running-ratio time--distance diagrams highlight the fast-propagating wave trains, which manifest themselves as narrow and steep stripes whose slopes represent the projection speeds of the wave trains on the plane of the sky. From these time--distance diagrams, one can see a long and broad stripe that represents the global EUV wave running ahead of the wave trains. The speed of the EUV wave along cut C1 is about 390 $\pm$ 10 \speed{}. It should be kept in mind that the propagation speed of the wave trains measured from time--distance diagrams are the lower limits of the true three-dimensional values due to projection effects. Although an obvious stationary brightening formed when the EUV wave reached a region of open magnetic fields, the EUV wave did not stop there but rather continued to propagate (see the black arrows in \fig{fig2}(b)), which may manifest the true wave nature of the EUV wave. In addition, by comparing the base-ratio time--distance diagrams, we can find that the initial global EUV front was followed by dimming in 171 \AA\ but emission enhancement in 193 \AA; this may suggest that the coronal structures were heated by the EUV wave through adiabatic heating \cite{schr11,down12,liu12}, and may not be due to the dissipation of the wave trains.

The wave trains have different manifestations in 171 \AA\ and 193 \AA\ time--distance diagrams. We mainly compare the time--distance diagrams made from 171 and 193 \AA\ running-ratio images along cut C2. In the 171 \AA\ time--distance diagram, we can observe the stripes of the wave trains at a distance of about 30 Mm from the measurement origin (see \fig{fig2}(e)), namely 150 Mm from the  footpoint of the guiding loop system. Before the wave trains interacted with the perpendicular loop system as indicated by the blue dash-dotted line \fig{fig2}(f), they propagated with an average speed (acceleration) of about 689 $\pm$ 23 \speed{} (\acc{-1043}).  However, this speed slowed down significantly to 343 $\pm$ 27 \speed{} after the interaction, about half of that before the interaction. This may indicate that the propagation of the wave trains was seriously influenced by the perpendicular loops due to the changing properties of the guiding loop system. In the 193 \AA\ running-ratio time--distance along the same cut (\fig{fig2}(f)), wave trains can only be identified after the interaction, and the stripes observed in 193 \AA\ time--distance diagram are weaker than those observed in the 171 \AA\ time--distance diagram. The average speed of the QFP wave trains measured from the 193 \AA\ time--distance diagrams is about 362 $\pm$ 36 \speed{}, while the acceleration is about \acc{-364}. This speed is slightly higher than that determined from the 171 \AA\ time--distance diagrams (\speed{343}), which may reflect the temperature response to the wave trains at different temperatures \cite{kidd12}.

\begin{figure}    
\centerline{\includegraphics[width=0.95\textwidth,clip=]{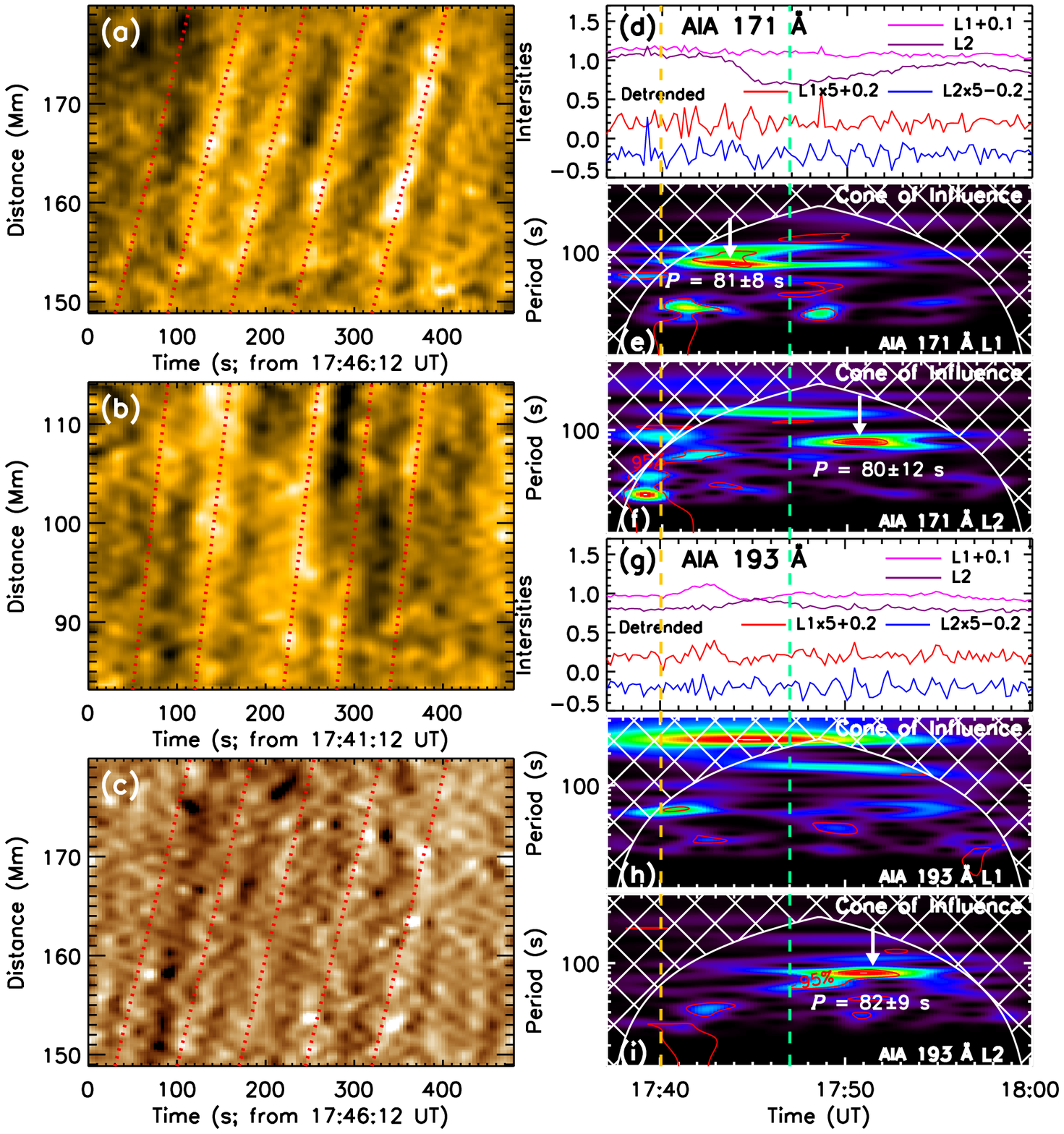}}
\caption{Periodicity analysis of the wave trains observed in 171 \AA\ and 193 \AA\ observations. Panels (a) and (b) display the close-up view of the top and bottom red dashed box regions shown in \fig{fig2}(e), while panel (c) is the region as shown in \fig{fig2}(f). In these time--distance diagrams, the QFP wave stripes are highlighted using a series of parallel dotted lines. In panel (d), the pink (magenta) curve shows the intensity profile along L1 (L2) as shown in \fig{fig2}(c), while the red (blue) curve displays the detrended intensity profile obtained by subtracting the smoothed flux using a 96 seconds boxcar. Panels (e) and (f) are the wavelet power spectra of the detrended intensity profiles along L1 and L2, respectively. Panels (g), (h), and (i) are to be compared with (d), (e), and (f), respectively, but they are for the 193 \AA\ intensity profiles. The red contours in each wavelet power spectrum outline the region where the significance level is above 95, \%, and the vertical yellow (green) line indicates the start time of the wave trains before (after) the interaction with the perpendicular loop system. In the power spectra, redder color correspond to higher wavelet power, and those with high power regions are indicated by vertical white arrows, and the corresponding periods ($\rm P$) are also plotted in the figure.}
\label{fig3}
\end{figure} 

\subsection{Periodic Analysis of the Wave Trains}
The detailed analysis of the periodicity of the wave trains is displayed in \fig{fig3}, in which panels (a) -- (c) are the magnified sub time--distance diagrams of the regions indicated by the red boxes shown in \fig{fig2}(e) and (f), while panels (d) -- (i) are wavelet analysis of the base-ratio intensity profiles along dashed lines L1 and L2, as shown in \fig{fig2}. In the sub time--distance diagrams, the steep stripes of the wave trains are clear and parallel to each other. We highlight these wave stripes using a series of parallel red dotted lines (see \fig{fig3}(a) -- (c)), and therefore, the time intervals between neighbouring lines represent the periods of the wave trains. The result indicates that the period of the QFP wave trains before their interaction with the perpendicular loop system ranges from 60 to 100 seconds (see \fig{fig3}(b)), while that ranges from 70 to 90 seconds after the interaction (see \fig{fig3}(a) and (c)). In addition, the QFP wave trains showed similar patterns and periods in the 171 \AA\ and 193 \AA\ time--distance diagrams after the interaction (see \fig{fig3} (a) and (c)), which suggests an intimate relationship between the wave trains observed at the two different wavelength bands. 

The base-ratio intensity profiles along L1 (pink) and L2 (magenta) of 171 \AA\ are plotted in \fig{fig3} (d), while those obtained from 193 \AA\ are plotted in \fig{fig3}(g). In optically thin corona, it is usually true that the emission intensity is proportional to the square of the plasma density, i.e. $I \propto \rho^{2}$. Thus the base-ratio intensity perturbations appropriately represent the variations of the plasma density relative to the pre-event background. To better show the intensity variations and the periodic patterns of the base-ratio intensity profiles, we also plot the detrended intensity profiles in \fig{fig3}(d) and (g) as shown by the red (L1) and blue curves (L2). The detrended intensity profiles are obtained by subtracting the smoothed intensities using a 96 second boxcar, and the results shown in the figure are fivefold magnifications of the original detrended profiles. To extract the periods of the wave trains, we apply wavelet analysis technique to the detrended intensity profiles along L1 and L2. The wavelet method is a common effective technique for analyzing localized variations of power within a time series, which allows us to investigate the temporal dependence periods within the observed data. The details of the procedure and the corresponding guidance can be found in \inlinecite{torr98}. In our analysis, we choose the ``Morlet'' function as the mother function, and a red-noise significance test is performed. Since both the time series and the wavelet function are finite, the wavelet can be altered by edge effects at the end of the time series. The significance of this edge effect is shown by a cone of influence (COI), defined as the region where the wavelet power drops by a factor of $\rm e^{-2}$. Areas of the wavelet power spectrum outside the region bounded by the COI should not be included in the analysis. 

The wavelet power spectra of 171 (193) \AA\ detrended intensity profiles along L1 and L2 are showed in \fig{fig3}(e) ((h)) and (f) ((i)) respectively. At the position L1, strong power with a period of $81 \pm 8$ seconds is identified. It starts from about 17:40 UT and lasts for about 12 minutes. However, no corresponding periodic signature could be detected at the same position in the 193 \AA\ intensity profile (see \fig{fig3}(h)). This is consistent with the imaging observations described above. At the position L2, we detect strong power with similar periods and durations both in the 171 and 193 \AA\ power spectra. The duration of this strong power is about eight minutes (17:47 UT -- 17:55 UT), and the periods are $80 \pm 12$ seconds and $82 \pm 9$ seconds in the 171 \AA\ and 193 \AA\ power spectra, respectively. At the two wavelength bands, the start times of the periodic signature are almost the same (see the vertical green line in \fig{fig3}). The similar periods revealed by the power spectra indicate that the wave trains kept their period before and after their interaction with the perpendicular loop system, even though their speed was slowed down significantly during the interaction. In our measurement, the periods are determined from the peak of the corresponding global power curve and meanwhile the significance level should be higher than 95, \%. The error of each period is determined by the full width at half maximum of each peak of the global power curve, which is obtained by fitting each peak with a Gaussian function.

\begin{figure}    
\centerline{\includegraphics[width=0.95\textwidth,clip=]{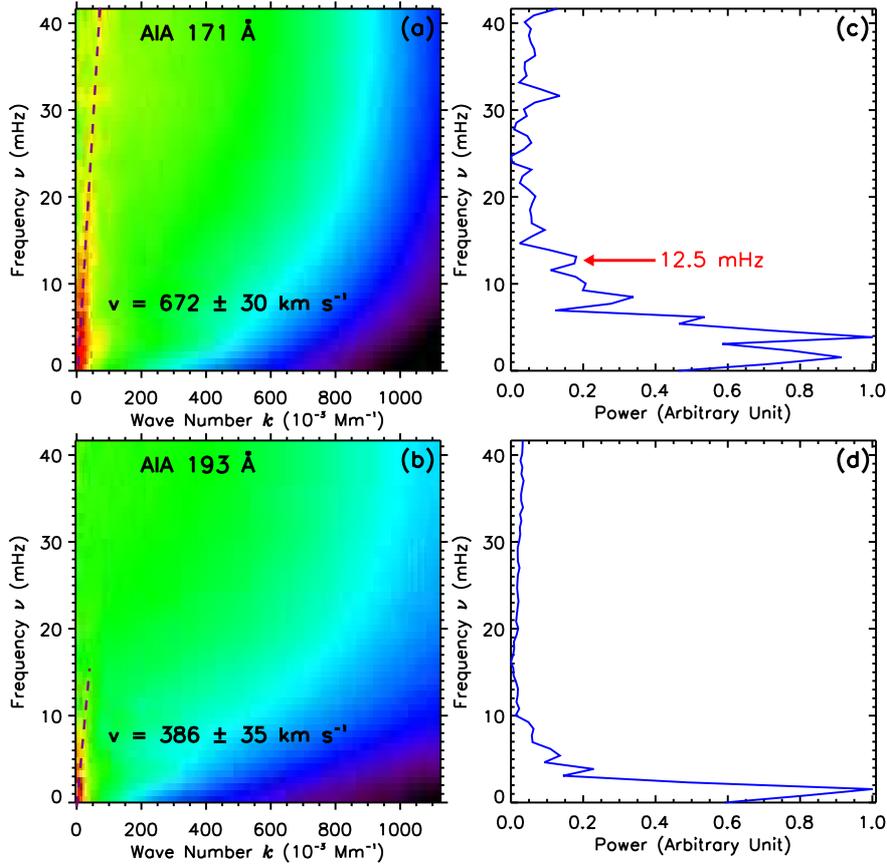}}
\caption{Fourier analysis of the QFP wave in the white dashed box region shown in Figure \ref{fig1}(b). Panels (a) and (b) are Fourier power ($k$--$\omega$ diagram) of a three-dimensional data tube of 171 \AA\ and 193 \AA\ running difference images during 17:40 -- 17:58 UT, while (c) and (d) are the integrated power spectrum over the wave number of left panels (a) and (b), respectively. The dashed line in panels (a) and (b) are the linear fit to the wave ridge.The red arrow in panel (c) point to the frequency of 12.5 mHz (period: 80 seconds).}
\label{fig4}
\end{figure}

To further analyze the periodicity of the wave trains observed in 171 \AA\ and 193 \AA\ observations, we generate $k$--$\omega$ diagrams from 171 \AA\ and 193 \AA\ running difference observations, with the Fourier transform method that can decompose the possible frequencies in the observed QFP wave. The principle and detailed operation steps have been documented in previous articles \cite{defo04, liu11, shen12b}. The analysis region is shown as the white dashed box in \fig{fig1}(b), and the analysis time is from 17:40 to 17:58 UT, close to the duration the QFP wave. The Fourier analysis results are shown in \fig{fig4}, in which panels (a) and (b) are the $k$--$\omega$ diagrams generated from 171 \AA\ and 193 \AA\ running-difference observations respectively. Based on the selected field of view of the analysis region and the temporal interval of the observation, we can obtain the resolution of the $k$--$\omega$ diagrams, which is $\Delta k= 6.85\times10^{-3}$ Mm$^{-1}$ in $x$-axis direction and $\Delta \nu=0.93$ mHz in $y$-axis. In each $k$--$\omega$ diagram, one can find an obvious linear steep ridge that represents the dispersion relation of the QFP wave, and it can be well fitted with a straight line passing through the origin (see the dashed lines in \fig{fig4} (a) and (b)). The slope of each ridge gives the phase speed [$v_{\rm ph}=\nu/k$] of the QFP wave, which is about 672 $\pm$ 30 \speed{} obtained from the 171 \AA\ $k$-$\omega$ diagram, while that is about 386 $\pm$ 35 \speed{} for the wave observed in 193 \AA\ observations. The speed revealed by the 193 \AA\ $k$-$\omega$ diagram is close to the average speed of the wave trains measured directly from the 193 \AA\ time--distance diagrams, whereas, the speed revealed by the 171 \AA\ $k$-$\omega$ diagram is just consistent with the average speed measured from the 171 \AA\ time--distance diagrams before the interaction with the perpendicular loop system. For each $k$--$\omega$ diagram, we plot the integrated power over the wave number in the right (see panels (c) and (d) in \fig{fig4}), which show a few peaks such as 1.3, 3.6 , 8.2, and 12.5 mHz for the 171 \AA\ Fourier power and 1.3, 3.5, and 5.1 for the 193 \AA. Among these frequencies, the frequency (period) 12.5 mHz (80 seconds) coincides with the period revealed by wavelet analysis of the intensity variations at the positions of L1 and L2, as well as the direct estimation from the time--distance diagrams in \fig{fig3}.

\begin{figure}    
\centerline{\includegraphics[width=0.95\textwidth,clip=]{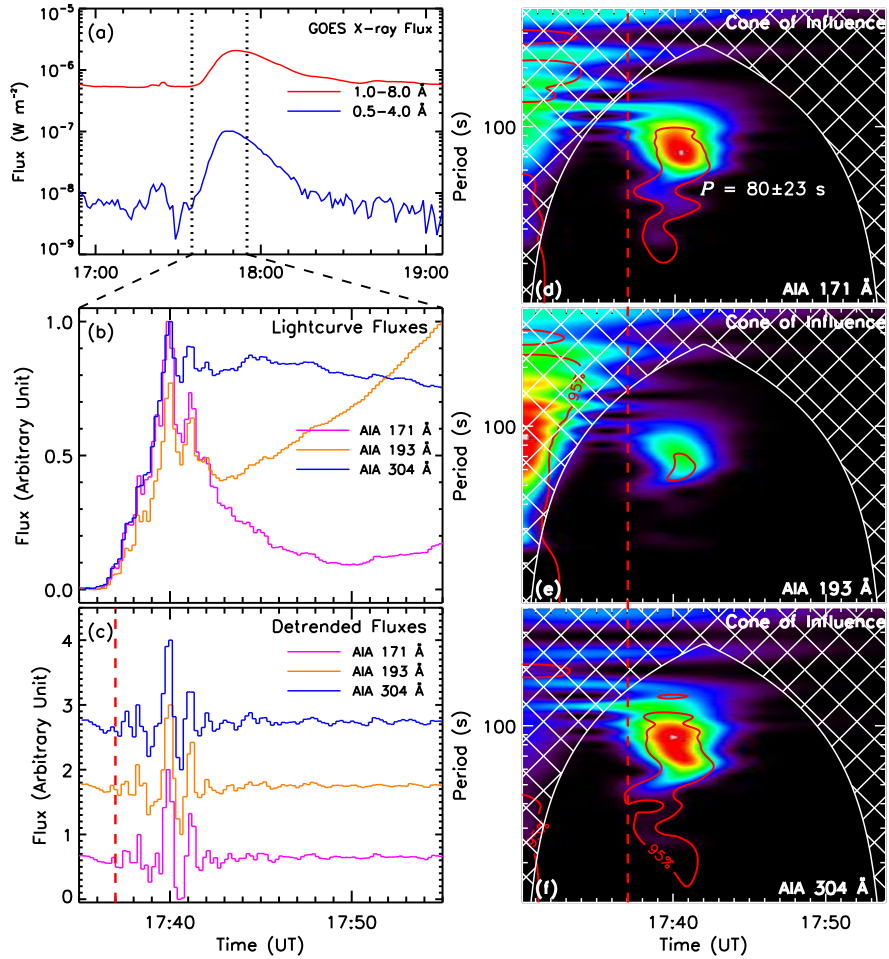}}
\caption{Periodicity analysis of the flare pulsations. Panel (a) is GOES soft X-ray fluxes, in which the red (blue) curve is the time profile of GOES 1 - 8 (0.5 - 4) \AA\ flux. Panel (b) shows the lightcurves of 171 \AA\  (pink), 193 \AA\ (yellow), and 304 \AA\ (blue) over the flare ribbon. The detrended 171 \AA\ (pink), 193 \AA\ (yellow), and 304 \AA\ (blue) fluxes are plotted in panel (c). (e) -- (f) are the wavelet power spectra of these detrended fluxes, in which the red contours indicate the region where the significance level is above 95, \%. The vertical red dashed lines in panels (c) -- (f) indicate the start time of the flare (17:37 UT).}
\label{fig5}
\end{figure}

\subsection{Periodic Analysis of the Flare Pulsation}
For impulsively launched fast waves in the low corona, flares are thought to be an obvious source \cite{asc05}. Recent high temporal and spatial resolution imaging results indicate that the associated flares have similar periods with the QFP waves (\opencite{liu11}; \citeyear{liu12}; \opencite{shen12b}). This may imply that the two phenomena are different manifestations of a single process such as magnetic reconnection. As expected, the QFP wave studied in this article shows an intimate relationship with the accompanying C2.0 flare. We use the light-curves over the flare ribbon close to the guiding loop's footpoint to analyze the periodicity of the flare pulsation. The GOES soft X-ray fluxes of 1.0 -- 8.0 \AA\ and 0.5 -- 4.0 \AA\ bands, flare light-curves of 171 \AA, 193 \AA, and 304 \AA, and the wavelet power of the corresponding detrended fluxes are show in \fig{fig5}. The two vertical dashed lines in \fig{fig5}(a) indicate a temporal interval from 17:35 UT to 17:55 UT, and the light-curves during this period are shown in panel (b). Panel (c) shows the detrended light-curves whose wavelet power spectra are shown in panels (d) -- (f). The detrended light-curves of 171 \AA, 193 \AA, and 304 \AA\ show coherence pulsations during the rising phase of the flare (see \fig{fig5}(c)). As it can be identified in the figure, the flare light-curves have a strong period of 80 seconds, in agreement with the period of the wave trains obtained by direct estimation from imaging observations. The similar period for both the flare and the wave trains implies that they were probably excited by a common physical origin, consisting with previous results \cite{liu11,shen12b}. In addition, the start time of the flare pulsation was the same as that of the flare, i.e. 17:37 UT, which is about three (tem) minutes earlier than the appearance time of the wave trains in the 171 (193) \AA\ observations.

\section{Discussions}
\subsection{The Generation of the QFP wave}
For impulsively generated fast magnetosonic waves in the solar corona, flares are thought to be an obvious source \cite{robe84,asc05}. However, up to the present, knowledge about the detailed generation mechanisms of the periodicity of flares and thereby QFP waves remain unclear, although previous studies, as well as the present study, have indicated that QFP waves have similar periods to the accompanying flares \cite{liu11,shen12b}. Based on these observational results, we propose that both QFP waves and the associated flares reflect the details of the energy releasing states in magnetic reconnections.

As summarized by \inlinecite{naka05b}, there are several physical mechanisms that can be responsible for flare periodicity, including i) geometrical resonances, ii) dispersive evolution of initially broadband signals, iii) nonlinear processes in magnetic reconnections, and iv) the leakage of oscillation modes from other layers of the solar atmosphere. For the present study, the last two mechanisms can be used to interpret the generation of the periodicity of the QFP wave. Since the period of 80 seconds can be identified in both the flare pulsation and the QFP wave, we propose that this component should be excited by some nonlinear processes in the magnetic reconnection that produces the flare. For example, recent numerical experiments indicate that repetitive generation of magnetic islands and their coalescence in current sheets are identified during magnetic reconnections, which can lead to an intermittent or impulsive bursty energy release \cite{klie00,mei12,ni12}. The generation of a new island is suggested to be accompanied by a burst of magnetic energy. The repetition of such process will form the periodicity of flares and QFP waves. In such a regime, the periods are determined by the properties of current sheet such as the plasma concentration, temperature, and magnetic field outside the current sheet \cite{naka09}. In addition, the so-called oscillatory reconnection could also be a possible mechanism for the generation of QFP waves (\inlinecite{murr09}; \inlinecite{mcla09}, \citeyear{mcla12}. Oscillatory reconnection releases energy periodically and thereby produce the repetitive pulsations of the flare emission. Up to the present, various mechanisms have been proposed to explain the periodicity of flare pulsation. However, which one is the corresponding mechanism for QFP wave still need to be proved.

Beside the common period of 80 seconds, a few low frequencies such as 1.3 ($P=770 seconds$) and 3.5 mHz ($P = 285 seconds$) are revealed by the $k$--$\omega$ diagrams of the QFP wave. These oscillation signatures are possibly the manifestations of the photospheric or chromospheric pressure-driven oscillations leaking into the solar corona. This mechanism has been identified in many observational and theoretical studies ({\it e.g.} \opencite{demo00}; \citeyear{demo02a}; \opencite{mars03}; \opencite{depo04}; \opencite{depo05}; \opencite{didk11}; \opencite{zaqa11}). Hence we can propose that the leakage of oscillation modes from the layers below the corona is also an important driving mechanism for the generation of the observed QFP wave in the low corona, in line with our previous results \cite{shen12b}.

\subsection{Propagation of the Wave Trains}
According to the observational results based on the 171 \AA\ observations, the propagation of the wave trains could be divided into three stages, the invisible stage (17:37 -- 17:40 UT), the fast propagation stage (\speed{689}), and the slow propagation stage (\speed{343}). The wave trains underwent an invisible stage of about three minutes before their appearance at a distance of about 150 Mm from the footpoint of the guiding loop system, during this stage no significant intensity perturbation could be observed. This may caused by the strong magnetic-field strength or other properties of the footpoint section of the guiding loop system, which may result in insufficient plasma compression and thereby no wave trains could be detected in the imaging observations. We can estimate the average speed during this stage by dividing the distance (150 Mm) by the length of time (180 seconds), which yields a speed of about \speed{833}. This result indicates that the speed during the fast propagation stage has been slowed down to about 80, \% of that during the invisible stage.

We can understand the deceleration of the wave trains from the basic equation of the fast magnetosonic wave when it propagates along magnetic field, i.e. $v_{f} =\frac{B}{\sqrt{4 \pi \rho}} (\theta = 0)$, $B$ being the magnetic field strength, $\rho$ the plasma density, and $\theta$ the angle between the guiding magnetic field and the wave vector \cite{asc05}. It can be seen that the propagation speed of the fast magnetosonic wave is determined by the magnetic field strength and the density of the medium that supports the wave. Considering the guiding loop system that has a divergent geometry and the gravitational stratification of the density with altitude, the speed of the fast magnetosonic wave would decrease rapidly with height due to the decrease of the magnetic-field strength with height \cite{ofma11}. In the meantime, if the total wave energy keeps unchanged during the propagation, the decrease of density with height will amplify the amplitudes of the wave trains and thereby become observable in the imaging observations. However, as the wave trains propagate outwards, the guiding loop become more and more diffuse. Therefore, the wave energy will spread to a broader extent, which will lead to the decrease of the amplitudes of the wave trains. The combined effects of the density stratification and the divergence geometry of the guiding loop can lead to the appearance of a maximum amplitude in the midway of the path as pointed out by \inlinecite{yuan13}. The quantitative relations among these parameters need to be investigate with numerical experiments.

After the wave trains interacted with the perpendicular loop system, the propagation entered a slow propagation stage with a speed of \speed{343} that is about half of that during the fast propagation stage. In the meantime, similar wave trains appeared in the 193 \AA\ observations, which has the same period and speed with those observed in the 171 \AA\ observations. The sudden decrease of the wave speed observed here could be interpreted from two aspects: the geometric effect and the density increase of the guiding loops. It is well known that the distribution of magnetic fields is very complex, but the basic configuration should be in funnel-like shape as proposed by \inlinecite{gabr76}. In the present case, the guiding loops carrying the wave trains may change their inclination angle significantly when approaching the perpendicular loop system, and thus the guiding loops become more curved upward over the underlaying perpendicular loops, i.e. a larger inclination angle relative to the solar surface. Therefore, due to projection effect the observed wave speed can decrease to a small value within a short timescale. Since the 193 \AA\ wavelength images higher layers of the solar corona than that of 171 \AA, and the wave trains propagated from a lower height from the footpoint of the guiding loops, the projection effect can also account for the sudden appearance of the wave trains in the 193 \AA\ observations. On the other hand, the sudden decrease of the wave speed can also be understood from the density increase of the guiding loop. When the wave-guiding loops interact with the underlying perpendicular loops, the wave trains will cause a strong compression of the guiding fields, which would increase the density of the guiding loops quickly and thereby decrease the speed of the wave trains within a short timescale. In addition, the compression can still cause a possible adiabatic heating that dissipates the wave energy and thus result in the wave trains in the 193 \AA\ observations.

\subsection{Estimation of Wave Energy and Magnetic Field}
We can measure the intensity variation of the wave trains in amplitude above the background with equation $I_{\rm A}=\frac{I_{\rm t}-I_{t0}}{I_{\rm t0}}$. The amplitude is determined from the wave crests and troughs during the prominent period of the wave trains. In 171 \AA\ observations, the amplitude variation along L1 is  2.3, \% - 5.0, \% of the background intensity, and the average value is 3.5, \% $\pm$ 0.8, \%. Along L2, the amplitude variations are 1.2, \% - 4.0, \% in 171 \AA\ and 0.3, \% - 3.7, \% of the background intensity in 193 \AA, and the average amplitude are 2.6, \% $\pm$ 1.1, \% (171 \AA) and 2.5, \% $\pm$ 1.3, \% (193 \AA) of the background intensity. The error of the average amplitude is given by the standard deviation of the measured values. It can be seen that the average amplitude of the wave trains weakened significantly after their interaction with the perpendicular loop system.

The energy flux carried by the QFP wave can be estimated from the kinetic energy of the perturbed plasma that propagate with phase speed through a volume element. So that the energy of the perturbed plasma is $E = (\frac{1}{2} \rho v_{\rm 1}^2) v_{\rm ph}$, where $v_{\rm 1}$ is the disturbance speed of the locally perturbed plasma \cite{asc04}, and $v_{\rm ph}$ is the phase speed. In the optically thin corona, it is usually true that $I \propto \rho ^{2}$. Thus the density modulation of the background density $\frac{\rm d \rho}{\rho} = \frac{\rm dI}{2I}$. In addition, if we use the relation $\frac{v_{\rm 1}}{v_{\rm ph}} \geq \frac{\rm d \rho}{\rho}$, then the energy flux of the perturbed plasma could be written as $E \geq \frac{1}{8} \rho v_{\rm ph}^{3} (\frac{\rm dI}{I})^{2}$ \cite{liu11}. For the present study, the average phase speeds during the fast, and slow propagation stages are \speed{689, and 343}, while the average amplitudes are 3.5, \% and 2.6, \% of the background intensity during the two stages respectvely. By assuming that the electron-number density of the wave guiding loops is $n_{\rm e} = 1 \times 10^{9}$ cm$^{-3}$, we can calculate that the energy-flux density of the QFP wave before and after the interaction are $E \geq 1.7 \times 10^{5}$ erg cm$^{-2}$ s$^{-1}$ and $E \geq 0.1 \times 10^{5}$ erg cm$^{-2}$ s$^{-1}$, respectively. As the typical energy-flux density requirement for heating coronal loops is about $10^{5}$ erg cm$^{-2}$ s$^{-1}$ \cite{with77,asc05}, so the energy flux carried by the QFP wave is sufficient for sustaining the coronal temperature of the guiding loops.
  
With the average speeds of the wave trains during the three distinct stages and the expression of the fast magnetosonic wave along magnetic fields [$v_{f} =\frac{B}{\sqrt{4 \pi \rho}}$ ($\theta = 0$)], we can estimate the magnetic-field strength of the guiding loops at different sections of the guiding loop system with $B=v_{f}\sqrt{4 \pi \rho}$. The calculation results indicate the magnetic-field strengths of the footpoint (invisible stage), middle (fast stage) and end (slow stage) sections of the guiding loop are 5.4, 4.5, and 2.2 Gauss, respectively. Since these values are calculated from the projection speeds, they are just the lower limits of the real magnetic-field strength values. In addition, we use the same density [$\rho$] in our calculation. Therefore, it should be kept in mind that the energy fluxes obtained and magnetic-field strength may only roughly reflect the true situation. Even so, the values obtained still reflect the distribution of magnetic-field strength along the divergence guiding loop system.

\section{Summary} 
With high temporal and spatial resolution observations taken by SDO/AIA, we present an observational study of a quasi-periodic fast-propagating magnetosonic wave along an open coronal loop rooted in active region AR11461. We study the generation, propagation, and the periodicity of the wave trains, as well as their relationship with the associated C2.0 flare. The wave trains first appeared in the 171 \AA\ observations at a distance of about 150 Mm from the footpoint of the guiding loops, then they were observed in the 193 \AA\ observations after their interaction with an underlying perpendicular loop system on the path. To our knowledge, such a phenomenon as well as multi-wavelength observations of QFP waves have not been studied in the past. The main observational results of the present study could be summarized as follow.

\begin{enumerate}
\item[i] The QFP wave trains and the associated flare have a common period of 80 seconds, which suggests that the generation of the wave trains and the flare pulsation originated from one common physical process. We propose that the periodic releasing of magnetic energy bursts through some regimes such as nonlinear processes in magnetic reconnections or the so-called oscillatory reconnections can account for the generation of the QFP wave trains. In addition, the component of the low frequencies revealed by the $k$--$\omega$  diagrams may be caused by the leakage of pressure-driven oscillations from the photosphere or chromosphere, which could be another important source for the generation of QFP waves in the low corona.
\item[ii] The propagation of the wave trains can be divided into three stages: the invisible stage (\speed{833}), the fast propagation stage (\speed{689}), and the slow propagation stage (\speed{343}). We conclude that the properties of the guiding loop have determined the manifestations of the wave trains during different stages, such as the distribution of the density and magnetic-field strength along the guiding loop system, as well as the geometry morphology.
\item[iii] The interaction of the wave trains with an underlaying perpendicular loop system is observed. This process caused two results: the sudden deceleration of the wave and the appearance of the wave trains in the 193 \AA\ observations. These phenomena are new observational results for QFP waves, and they can be understood from the geometric effect and the density increase of the guiding loop system due to the interaction between the wave trains and the underlying perpendicular loop system. The interaction may also have caused the heating of the cool plasma to higher temperature through adiabatic heating.
\item[iv] The amplitude of the wave trains is measured. In the 171 \AA\ observations, the average values is about 3.5, \% (2.6, \%) of the background intensity before (after) the interaction with the perpendicular loop system, and that is about 2.5, \% in the 193 \AA\ observations. Based on these results, we estimate the energy flux density of the QFP wave and the magnetic-field strength of the guiding loop system. The order of magnitude of the energy flux carried by the QFP wave is of $10^{5}$ erg cm$^{-2}$ s$^{-1}$, which is sufficient to sustaining the coronal temperature of the guiding loops. The magnetic-field strength estimated from the wave speeds indicates the distribution of the divergence geometry of the wave guiding loops. From the footpoint of the guiding loops to the other end, the estimated mean magnetic-field strength decreases from 5.4 to 2.2 Gauss.
\end{enumerate}

In summary, the interesting QFP waves could be used for remote diagnostics of the local physical properties of the solar corona. However, details about the generation, propagation and energy dissipation of QFP waves are still unclear. Further theoretical and statistical studies on QFP waves are required.

\begin{acks}
The authors thank the data support of GOES and SDO which is a mission for NASA's Living With a Star (LWS) program. We thank an anonymous referee for many helpful comments and valuable for improving the quality of this article. The wavelet software is provided by C. Torrence and G. Compo. It is available at \urlurl{atoc.colorado.edu/research/wavelets}. This work is supported by the Natural Science Foundation of China under grants 10933003, 11078004, and 11073050, the Ntional Key Research Science Foundation (2011CB811400), the Knowledge Innovation Program of the CAS (KJCX2-EW-T07), the Western Light Youth Project of CAS, the Open Research Program of the Key Laboratory of Solar Activity of CAS (KLSA201204, KLSA201219), and the open research program of the Key Laboratory of Dark Matter and Space Astronomy of CAS (DMS2012KT008). A. Elmhamdi is supported by the CAS fellowships for young international scientists under grant number 2012Y1JA0002.
\end{acks}

\end{article} 

\begin{thebibliography}{}
\bibitem[\protect\citeauthoryear{{Appert \etal}}{1986}]{appe86}
Appert, K., Collins, G.A., Hellsten, T., Vaclavik, J., Villard, L.: 1986, {\it Plasma Phys. and Control. Fusion}{} \textbf{28}, 133.
\bibitem[\protect\citeauthoryear{{Aschwanden}}{2004}]{asc04}
Aschwanden, M.J.: 2004, In: Coronal Heating, eds. Walsh R. W., Ireland J., Danesy D., Fleck B. (SP-575, ESA), 97.
\bibitem[\protect\citeauthoryear{{Aschwanden}}{2005}]{asc05}
Aschwanden, M.J.: 2005, Physics of the Solar Corona. An Introduction with Problems and Solutions, 2nd ed.; Praxis Publishing Ltd., Chichester.
\bibitem[\protect\citeauthoryear{{Boerner \etal}}{2012}]{boer12}
Boerner, P., Edwards, C., Lemen, J., Rausch, A., Schrijver, C., Shine, R., \etal: 2012, \solphys{} \textbf{275}, 41.
\bibitem[\protect\citeauthoryear{{Bogdan \etal}}{2003}]{bogd03}
Bogdan, T.J., Carlsson, M., Hansteen, V., McMurry, A., Rosenthal, C.S., Johnson, M., \etal: 2003, \apj{} \textbf{599}, 626.
\bibitem[\protect\citeauthoryear{{DeForest}}{2004}]{defo04}
DeForest, C.E.: 2004, \apjl{} \textbf{617}, L89.
\bibitem[\protect\citeauthoryear{{De Moortel, Ireland, and Walsh}}{2000}]{demo00}
De Moortel, I., Ireland, J., Walsh, R.W.: 2000, \aap{} \textbf{355}, L23.
\bibitem[\protect\citeauthoryear{{De Moortel, Ireland, and Walsh}}{2002}]{demo02a}
De Moortel, I., Ireland, J., Hood, A.W., Walsh, R.W.: 2002, \aap{} \textbf{387}, L13.
\bibitem[\protect\citeauthoryear{{De Pontieu, Erd$\rm \acute{e}$lyi, and De Moortel}}{2005}]{depo05}
De Pontieu, B., Erd$\rm \acute{e}$lyi, R., De Moortel, I.: 2005, \apjl{} \textbf{624}, L61.
\bibitem[\protect\citeauthoryear{{De Pontieu, Erd$\rm \acute{e}$lyi, and James}}{2004}]{depo04}
De Pontieu, B., Erd$\rm \acute{e}$lyi, R., James, S.P.: 2004, \nat{} \textbf{430}, 536.
\bibitem[\protect\citeauthoryear{{Didkovsky \etal}}{2011}]{didk11}
Didkovsky, L., Judge, D., Kosovichev, A.G., Wieman, S., Woods, T.: 2011, \apjl{} \textbf{738}, L7.
\bibitem[\protect\citeauthoryear{{Downs \etal}}{2012}]{down12}
Downs, C., Roussev, I.I., van der Holst, B., Lugza, N., Sokolov, I. V.: 2012, \apj{} \textbf{750}, 134.
\bibitem[\protect\citeauthoryear{{Edwin and Roberts}}{1983}]{edwi83}
Edwin, P.M., Roberts, B.: 1983, \solphys{} \textbf{88}, 179.
\bibitem[\protect\citeauthoryear{{Edwin and Roberts}}{1988}]{edwi88}
Edwin, P.M., Roberts, B.: 1988, \aap{} \textbf{192}, 343.
\bibitem[\protect\citeauthoryear{{Fedun, Shelyag, and Erd$\rm \acute{e}$lyi}}{2011}]{fedu11}
Fedun, V., Shelyag, S., Erd$\rm \acute{e}$lyi, R.: 2011, \apj{} \textbf{727}, 17.
\bibitem[\protect\citeauthoryear{{Gabriel}}{1976}]{gabr76}
Gabriel, A.H.: 1976, {\sl Phil. Trans. Roy. Soc. Ser. A}{} \textbf{281}, 339.
\bibitem[\protect\citeauthoryear{{Heggland, De Pontieu, and Hansteen}}{2009}]{hegg09}
Heggland, L., De Pontieu, B., Hansteen, V.H.: 2009, \apj{} \textbf{702}, 1.
\bibitem[\protect\citeauthoryear{{Jel\'{i}nek, Karlick\'{y}, and Murawski}}{2012}]{jeli12}
Jel\'{i}nek, P., Karlick\'{y}, M., Murawski, K.: 2012, \aap{} \textbf{546}, A49.
\bibitem[\protect\citeauthoryear{{Jiang \etal}}{2007}]{jian07}
Jiang, Y.C., Chen, H.D., Shen, Y.D., Yang, L.H., Li, K.J.: 2007, \solphys{} \textbf{240}, 77.
\bibitem[\protect\citeauthoryear{{Karlick\'{y}, Jel\'{i}nek, P., and M\'{e}sz\'{a}rosov\'{a}}}{2011}]{karl11}
Karlick\'{y}, M., Jel\'{i}nek, P., M\'{e}sz\'{a}rosov\'{a}, H.: 2011, \aap{} \textbf{529}, A96.
\bibitem[\protect\citeauthoryear{{Kiddie \etal}}{2012}]{kidd12}
Kiddie, G., De Moortel, I., Del Zanna, G., McIntosh, S.W., Whittaker, I.: 2012, \solphys{} \textbf{279}, 427.
\bibitem[\protect\citeauthoryear{{King \etal}}{2003}]{king03}
King, D.B., Nakariakov, V.M., DeLuca, E.E., Golub, L., McClements, K.G.: 2003, \aap{} \textbf{404}, L1.
\bibitem[\protect\citeauthoryear{{Kliem, Karlick$\rm \acute{y}$, and Benz}}{2000}]{klie00}
Kliem, B., Karlick$\rm \acute{y}$, M., Benz, A.O.: 2000, \aap{} \textbf{360}, 715.
\bibitem[\protect\citeauthoryear{{Koutchmy, $\rm \check{Z}$ug$\rm \check{z}$da, and Loc$\rm \check{a}$ns}}{1983}]{kout83}
Koutchmy, S., $\rm \check{Z}$ug$\rm \check{z}$da, Y.D., Loc$\rm \check{a}$ns, V.: 1983, \aap{} \textbf{120}, 185.
\bibitem[\protect\citeauthoryear{{Lemen \etal}}{2012}]{leme12}
Lemen, J.R., Title, A.M., Akin, D.J., Boerner, P.F., Chou, C., Drake, J.F., \etal: 2012, \solphys{} \textbf{275}, 17.
\bibitem[\protect\citeauthoryear{{Liu \etal}}{2012}]{liu12}
Liu, W., Ofman, L., Nitta, N.V., Aschwanden, M.J., Schrijver, C.J., Title, A.M., \etal: 2012, \apj{} \textbf{753}, 52.
\bibitem[\protect\citeauthoryear{{Liu \etal}}{2011}]{liu11}
Liu, W., Title, A.M., Zhao, J., Ofman, L., Schrijver, C.J., Aschwanden, M.J., \etal:  2011, \apjl{} \textbf{736}, L13.
\bibitem[\protect\citeauthoryear{{Marsh \etal}}{2003}]{mars03}
Marsh, M. S., Walsh, R. W., De Moortel, I., Ireland, J.: 2003, \aap{} \textbf{404}, L37.
\bibitem[\protect\citeauthoryear{{McLaughlin \etal}}{2009}]{mcla09}
McLaughlin, J.A., De Moortel, I., Hood, A.W., Brady, C.S.: 2009, \aap{} \textbf{493}, 227.
\bibitem[\protect\citeauthoryear{{McLaughlin \etal}}{2012}]{mcla12}
McLaughlin, J.A., Verth, G., Fedun, V., Erd\'{e}lyi, R.: 2012, \apj{} \textbf{749}, 30.
\bibitem[\protect\citeauthoryear{{Mei \etal}}{2012}]{mei12}
Mei, Z., Shen, C., Wu, N., Lin, J., Murphy, N.A., Roussev, I.I.: 2012, \mnras{} \textbf{425}, 2824.
\bibitem[\protect\citeauthoryear{{M\'{e}sz\'{a}rosov\'{a}, Karlick\'{y}, and Ryb\'{a}k}}{2011}]{mesz11}
M\'{e}sz\'{a}rosov\'{a}, H., Karlick\'{y}, M., Ryb\'{a}k, J.: 2011, \solphys{} \textbf{273}, 393.
\bibitem[\protect\citeauthoryear{{M\'{e}sz\'{a}rosov\'{a} \etal}}{2009}]{mesz09a}
M\'{e}sz\'{a}rosov\'{a}, H., Karlick\'{y}, M., Ryb\'{a}k, J., Ji\v{r}i\v{c}ka, K.: 2009a, \apj{} \textbf{697}, L108.
\bibitem[\protect\citeauthoryear{{M\'{e}sz\'{a}rosov\'{a} \etal}}{2009}]{mesz09b}
M\'{e}sz\'{a}rosov\'{a}, H., Karlick\'{y}, M., Ryb\'{a}k, J., Ji\v{r}i\v{c}ka, K.: 2009b, \aap{} \textbf{502}, L13.
\bibitem[\protect\citeauthoryear{{M\'{e}sz\'{a}rosov\'{a} \etal}}{2013}]{mesz13}
M\'{e}sz\'{a}rosov\'{a}, H., Dud\'{i}k, J., Karlick\'{y}, M., Madsen, F.R.H., Sawant, H.S.: 2013, \solphys{} \textbf{283}, 473.
\bibitem[\protect\citeauthoryear{{Murray, Driel-Gesztelyi, and Baker}}{2009}]{murr09}
Murray, M.J, van Driel-Gesztelyi, L., Baker, D.: 2009, \aap{} \textbf{494}, 329.
\bibitem[\protect\citeauthoryear{{Morton \etal}}{2012}]{mort12a}
Morton, R.J., Verth, G., Jess, D.B., Kuridze, D., Ruderman, M.S., Mathioudakis, M., \etal: {\sl Nature Communications}{} \textbf{3}, 1315.
\bibitem[\protect\citeauthoryear{{Morton \etal}}{2012}]{mort12b}
Morton, R.J., Verth, G., McLaughlin, J.A., Erd\'{e}lyi, R.: \apj{} \textbf{744}, 5.
\bibitem[\protect\citeauthoryear{{Nakariakov and Melnikov}}{2009}]{naka09}
Nakariakov, V.M., Melnikov, V.F.: 2009, \ssr{} \textbf{149}, 119.
\bibitem[\protect\citeauthoryear{{Nakariakov and Ofman}}{2001}]{naka01}
Nakariakov, V.M., Ofman, L.: 2001, \aap{} \textbf{372}, L53.
\bibitem[\protect\citeauthoryear{{Nakariakov and Verwichte}}{2005}]{naka05}
Nakariakov, V.M., Verwichte, E.: 2005, {\it Living Rev. Sol. Phys.}{} \textbf{2}, 3.
\bibitem[\protect\citeauthoryear{{Nakariakov, Melnikov, and Reznikova}}{2003}]{naka03}
Nakariakov, V.M., Melnikov, V.F., Reznikova, V.E.: 2003, \aap{} \textbf{412}, L7.
\bibitem[\protect\citeauthoryear{{Nakariakov, Pascoe, and Arber}}{2005}]{naka05b}
Nakariakov, V.M., Pascoe, D. J., Arber, T.D.: 2005, \ssr{} \textbf{121}, 115.
\bibitem[\protect\citeauthoryear{{Nakariakov \etal}}{1999}]{naka99}
Nakariakov, V.M., Ofman, L., Deluca, E.E., Roberts, B., Davila, J.M.: 1999, {\it Science}{} \textbf{285}, 862
\bibitem[\protect\citeauthoryear{{Nakariakov \etal}}{2004}]{naka04}
Nakariakov, V.M., Arber, T.D., Ault, C.E., Katsiyannis, A.C., Williams, D.R., Keenan, F.P.: 2004, \mnras{} \textbf{349}, 705.
\bibitem[\protect\citeauthoryear{{Ni \etal}}{2012}]{ni12}
Ni, L., Roussev, I.I., Lin, J., Ziegler, U.: 2012, \apj{} \textbf{758}, 20.
\bibitem[\protect\citeauthoryear{{Ofman \etal}}{2011}]{ofma11}
Ofman, L., Liu, W., Title, A., Aschwanden, M.: 2011, \apjl{} \textbf{740}, L33.
\bibitem[\protect\citeauthoryear{{Osterbrock}}{1961}]{oste61}
Osterbrock, D.: 1961, \apj{} \textbf{134}, 347.
\bibitem[\protect\citeauthoryear{{Parks and Winckler}}{1969}]{park69}
Parks, G.K., Winckler, J.R.: 1969, \apjl{} \textbf{155}, L117.
\bibitem[\protect\citeauthoryear{{Pesnell, Thompson, and Chamberlin}}{2012}]{pesn12}
Pesnell, W.D., Thompson, B.J., Chamberlin, P.C.: 2012, \solphys{} \textbf{275}, 3.
\bibitem[\protect\citeauthoryear{{Robbrecht \etal}}{2001}]{robb01}
Robbrecht, E., Verwichte, E., Berghmans, D., Hochedez, J.F., Poedts, S., Nakariakov, V.M.: 2001, \aap{} \textbf{370}, 591.
\bibitem[\protect\citeauthoryear{{Roberts, Edwin, and Benz}}{1983}]{robe83}
Roberts, B., Edwin, P.M., Benz, A.Q.: 1983, \nat{} \textbf{305}, 688.
\bibitem[\protect\citeauthoryear{{Roberts, Edwin, and Benz}}{1984}]{robe84}
Roberts, B., Edwin, P.M., Benz, A.Q.: 1984, \apj{} \textbf{279}, 857.
\bibitem[\protect\citeauthoryear{{Schatzman}}{1949}]{scha49}
Schatzman, E.: {\sl Annal. d'Astrophys.}{} \textbf{12}, 203.
\bibitem[\protect\citeauthoryear{{Schrijver \etal}}{2011}]{schr11}
Schrijver, C.J., Aulanier, G., Title, A.M., Pariat, E., Delann\'{e}e, C.: 2011, \apj{} \textbf{738}, 167.
\bibitem[\protect\citeauthoryear{{Shen and Liu}}{2012a}]{shen12b}
Shen, Y.D., Liu, Y.: 2012a, \apj{} \textbf{753}, 53.
\bibitem[\protect\citeauthoryear{{Shen and Liu}}{2012b}]{shen12c}
Shen, Y.D., Liu, Y.: 2012b, \apj{} \textbf{754}, 7.
\bibitem[\protect\citeauthoryear{{Shen and Liu}}{2012c}]{shen12a}
Shen, Y.D., Liu, Y.: 2012c, \apjl{} \textbf{752}, L23.
\bibitem[\protect\citeauthoryear{{Shen \etal}}{2013}]{shen13}
Shen, Y.D., Liu, Y., Su, J.T., Li, H., Zhao, R.J., Tian, Z.J., \etal: 2013, \apj{} \textbf{773}, L33.
\bibitem[\protect\citeauthoryear{{Shen \etal}}{2010}]{shen10}
Shen, Y.D., Li, K.J., Yang, L.H., Yang, J.Y., Jiang, Y.C.: {\sl Acta Astron Sinica}{} \textbf{51}, 151.
\bibitem[\protect\citeauthoryear{{Thompson \etal}}{1998}]{thom98}
Thompson, B.J., Plunkett, S.P., Gurman, J.B., Newmark, J.S., St. Cyr, O.C., Michels, D.J.: 1998, \jgr{} \textbf{25}, 2465.
\bibitem[\protect\citeauthoryear{{Tian, McIntosh, and De Pontieu}}{2011}]{tian11}
Tian, H., McIntosh, S.W., De Pontieu, B.: 2011, \apjl{} \textbf{727}, L37.
\bibitem[\protect\citeauthoryear{{Torrence and Compo}}{1998}]{torr98}
Torrence, C., Compo, G.P.: 1998, {\it Bull. Meteorol. Soc.}{} \textbf{79}, 61.
\bibitem[\protect\citeauthoryear{{Uchida}}{1970}]{uchi70}
Uchida, Y.: 1970, \pasj{} \textbf{22}, 341.
\bibitem[\protect\citeauthoryear{{Verwichte, Nakariakov, and Cooper}}{2005}]{verw05}
Verwichte, E., Nakariakov, V.M., Cooper, F.F.: 2005, \aap{} \textbf{430}, L65.
\bibitem[\protect\citeauthoryear{{Walsh and Ireland}}{2003}]{wals03}
Walsh, R.W., Ireland, J.: 2003, {\sl Astron. Astrophys. Rev.}{} \textbf{12}, 1.
\bibitem[\protect\citeauthoryear{{West \etal}}{2011}]{west11}
West, M.J., Zhukov, A.N., Dolla, L., Rodriguez, L.: 2011, \apj{} \textbf{730}, 122.
\bibitem[\protect\citeauthoryear{{Williams \etal}}{2001}]{will01}
Williams, D.R., Phillips, K.J.H., Rudawy, P., Mathioudakis, M., Gallagher, P.T., O'Shea, E., \etal: 2001, \mnras{} \textbf{326}, 428.
\bibitem[\protect\citeauthoryear{{Williams \etal}}{2002}]{will02}
Williams, D.R., Mathioudakis, M., Gallagher, P.T., Phillips, K.J.H., M$^{c}$Ateer, R.T.J., Keenan, F.P., \etal: 2002, \mnras{} \textbf{336}, 747.
\bibitem[\protect\citeauthoryear{{Withbroe and Noyes}}{1977}]{with77}
Withbroe, G.L., Noyes, R.W.: 1977, {\it Ann. Rev. Astron. Astrophys.}{} \textbf{15}, 363.
\bibitem[\protect\citeauthoryear{{Yuan \etal}}{2013}]{yuan13}
Yuan, D., Shen, Y., Liu, Y., Nakariakov, V. M., Tan, B., Huang, J.: 2013, \aap{} \textbf{554}, A144.
\bibitem[\protect\citeauthoryear{{Zaqarashvili \etal}}{2011}]{zaqa11}
Zaqarashvili, T.V., Murawski, K., Khodachenko, M.L., Lee, D.: 2011, \aap{} \textbf{529}, A85.
\end{thebibliography}
\end{document}